\documentclass[sigconf]{acmart}

\usepackage{booktabs} 

\usepackage{amsthm}
 
\setcopyright{acmcopyright}

\usepackage{hyperref}
\usepackage{ctable}
\hypersetup{
	colorlinks   = true,
       linkcolor = red,
            urlcolor  = blue,
            citecolor = blue
}
\usepackage{url}

\begin{document}
\title{Extracting Hierarchies of Search Tasks \& Subtasks via a Bayesian Nonparametric Approach}

\author{Rishabh Mehrotra$^{\dagger}$ and Emine Yilmaz$^{\dagger *}$}
\affiliation{
  \institution{$^{\dagger}$University College London, London, United Kingdom}
}
\affiliation{
  \institution{$^*$The Alan Turing Institute, British Library, London, United Kingdom}
}
\email{{r.mehrotra, e.yilmaz}@cs.ucl.ac.uk}

\copyrightyear{2017} 
\acmYear{2017} 
\setcopyright{acmcopyright}
\acmConference{SIGIR '17}{August 07-11}{2017, Shinjuku, Tokyo, Japan}
\acmPrice{15.00}\acmDOI{10.1145/3077136.3080823}
\acmISBN{978-1-4503-5022-8/17/08}

\renewcommand{\shorttitle}{Extracting Hierarchies of Search Tasks \& Subtasks}


\begin{abstract}
A significant amount of search queries originate from some real world information need or tasks~\cite{jones2008beyond}. In order to improve the search experience of the end users, it is important to have accurate representations of tasks. As a result, significant amount of research has been devoted to extracting proper representations of tasks in order to enable search systems to help users complete their tasks, as well as providing the end user with better query suggestions~\cite{hassan2014supporting}, for better recommendations~\cite{zhang2015task}, for satisfaction prediction~\cite{wang2014modeling} and for improved personalization in terms of tasks~\cite{mehrotra2015terms,white2013enhancing}. Most existing task extraction methodologies focus on representing tasks as flat structures. However, tasks often tend to have multiple subtasks associated with them and a more naturalistic representation of tasks would be in terms of a hierarchy, where each task can be composed of multiple (sub)tasks. To this end, we propose an efficient Bayesian nonparametric model for extracting hierarchies of such tasks \& subtasks. We evaluate our method based on real world query log data both through quantitative and crowdsourced experiments and highlight the importance of considering task/subtask hierarchies.
\end{abstract}


\maketitle
\section{Introduction}
The need for search often arises from a person's need to achieve a goal, or a task such as booking travels, buying a house, etc., which would lead to search processes that are often lengthy, iterative, and are characterized by distinct stages and shifting goals. ~\cite{jones2008beyond}. Thus, identifying and representing these tasks properly is highly important for devising search systems that can help end users complete their tasks. It has previously been shown that these task representations can be used to provide users with better query suggestions \cite{hassan2014supporting}, offer improved personalization \cite{mehrotra2015terms,white2013enhancing}, provide better recommendations \cite{zhang2015task}, help in satisfaction prediction \cite{wang2014modeling} and search result re-ranking. Moreover, accurate representations of tasks could also be highly useful in aptly placing the user in the task-subtask space to contextually target the user in terms of better recommendations and advertisements, developing task specific ranking of documents, and developing task based evaluation metrics to model user satisfaction. Given the wide range of applications these tasks representations can be used for, significant amount of research has been devoted to task extraction and representation \cite{lucchese2013discovering,hua2013identifying,kotov2011modeling,jones2008beyond,li2014identifying}.

Task extraction is quite a challenging problem as search engines can be used to achieve very different tasks, and each task can be defined at different levels of granularity. A major limitation in existing task-extraction methods lies in their treatment of search tasks as flat structure-less clusters which inherently lack insights about the presence or demarcation of subtasks associated with individual search tasks. In reality, often search tasks tend to be hierarchical in nature. For example, a search task like planning a wedding involves subtasks like searching for dresses, browsing different hairstyles, looking for invitation card templates, finding planners, among others. Each of these subtasks (1) could themselves be composed of multiple subtasks, and (2) would warrant issuing different queries by users to accomplish them. Hence, in order to obtain more accurate representations of tasks, new methodologies for constructing hierarchies of tasks are needed.

As part of the proposed research, we consider the challenge of extracting hierarchies of search tasks and their associated subtasks from a search log given just the log data without the need of any manual annotation of any sort. In a recent poster we showed that Bayesian nonparametrics have the potential to extract a hierarchical representation of tasks~\cite{mehrotra2015towards}; we extend this model further to form more accurate representations of tasks.

We present an efficient Bayesian nonparametric model for discovering hierarchies and propose a tree based nonparametric model to discover this rich hierarchical structure of tasks/subtasks embedded in search logs. Most existing hierarchical clustering techniques result in binary tree structures with each node decomposed into two child nodes. Given that a complex task could be composed of an arbitrary number of subtasks, these techniques cannot directly be used to construct accurate representations of tasks. In contrast, our model is capable of identifying task structures that can be composed of an arbitrary number of children. We make use of a number of evaluation methodologies to evaluate the efficacy of the proposed task extraction methodology, including quantitative and qualitative analyses along with crowdsourced judgment studies specifically catered to evaluating the quality of the extracted task hierarchies. We contend that the techniques presented expand the scope for better recommendations and search personalization and opens up new avenues for recommendations specifically targeting users based on the tasks they involve in.

\section{Related Work}
Web search logs provide explicit clues about the information seeking behavior of users and have been extensively studied to improve search experiences of users. We cover several areas of related work and discuss how our work relates to and extends prior work.

\subsection{Task Extraction}
There has been a large body of work focused on the problem of segmenting and organizing query logs into semantically coherent structures. Many such methods use the idea of a \textit{timeout} cutoff between queries, where two consecutive queries are considered as two different sessions or tasks if the time interval between them exceeds a certain threshold \cite{catledge1995characterizing,he2002combining,silverstein1999analysis}. Often a 30-minute timeout is used to segment sessions. 
However, experimental results of these methods indicate that the timeouts are of limited utility in predicting whether two queries belong to the same task, and unsuitable for identifying session boundaries.

More recent studies suggest that users often seek to complete multiple search tasks within a single search session \cite{mehrotra2016characterizing,lucchese2011identifying} with over 50\% of search sessions having more than 2 tasks \cite{mehrotra2016characterizing}. At the same time, certain tasks require significantly more effort, time and sessions to complete with almost 60\% of complex information gathering tasks continued across sessions \cite{agichtein2012search,ma2008exploring}. There have been attempts to extract in-session tasks \cite{jones2008beyond,lucchese2011identifying,spink2005multitasking}, and cross-session tasks \cite{kotov2011modeling,wang2013learning} from query sequences based on classification and clustering methods, as well as supporting users in accomplishing these tasks \cite{hassan2014supporting}. Prior work on identifying search-tasks focuses on task extraction from search sessions with the objective of segmenting a search session into disjoint sets of queries where each set represents a different task \cite{lucchese2013discovering,hua2013identifying}.

Kotov et al. \cite{kotov2011modeling} and Agichtein et al. \cite{agichtein2012search} studied the problem of cross-session task extraction via binary same-task classification, and found different types of tasks demonstrate different life spans. While such task extraction methods are good at linking a new query to an on-going task, often these query links form long chains which result in a task cluster containing queries from many potentially different tasks. With the realization that sessions are not enough to represent tasks, recent work has started exploring cross-section task extraction, which often results in complex non-homogeneous clusters of queries solving a number of related yet different tasks. Unfortunately, pairwise predictions alone cannot generate the partition of tasks efficiently and even with post-processing, the final task partitions obtained are not expressive enough to demarcate subtasks \cite{liao2012evaluating}. Finally, authors in \cite{li2014identifying} model query temporal patterns using a special class of point process called Hawkes processes, and combine topic model with Hawkes processes for simultaneously identifying and labeling search tasks.

Jones et al. \cite{jones2008beyond} was the first work to consider the fact that there may be multiple subtasks associated with a user's information need and that these subtasks could be interleaved across different sessions. However, their method only focuses on the queries submitted by a single user and attempts to segment them based on whether they fall under the same information need. Hence, they only consider solving the task boundary identification and same task identification problem and cannot be used directly for task extraction. Our work alleviates the same user assumption and considers queries across different users for task extraction. Finally, in a recent poster \cite{mehrotra2015towards}, we proposed the idea of extracting task hierarchies and presented a basic tree extraction algorithm. Our current work extends the preliminary model in a number of dimensions including novel model of query affinities and task coherence based pruning strategy, which we observe gives substantial improvement in results. Unlike past work, we also present detailed derivation and evaluation of the extracted hierarchy and application on task extraction.

\subsection{Supporting Complex Search Tasks}
There has been a significant amount of work on task continuation assistance \cite{morris2008searchbar,agichtein2012search}, building task tours and trails \cite{o2010tweetmotif,singla2010studying}, query suggestions \cite{baeza2005query,jones2006generating,mei2008query}, predicting next search action \cite{cao2009towards} and notes taking when accomplishing complex tasks \cite{donato2010you}. The quality of most of these methods depends on forming accurate representations of tasks, which is the problem we are addressing in this paper.

\subsection{Hierarchical Models}
Rich hierarchies are common in data across many domains, hence quite a few hierarchical clustering techniques have been proposed. The traditional methods for hierarchically clustering data are bottom-up agglomerative algorithms. Probabilistic methods of learning hierarchies have also been proposed \cite{blundell2013bayesian,liu2012automatic} along with hierarchical clustering based methods \cite{heller2005bayesian,chuang2002towards}. Most algorithms for hierarchical clustering construct binary tree representations of data, where leaf nodes correspond to data points and internal nodes correspond to clusters. There are several limitations to existing hierarchy construction algorithms. The algorithms provide no guide to choosing the correct number of clusters or the level at which to prune the tree. It is often difficult to know which distance metric to choose. Additionally and more importantly, restriction of the hypothesis space to binary trees alone is undesirable in many situations - indeed, a task can have any number of subtasks, not necessarily two. Past work has also considered constructing task-specific taxonomies from document collections \cite{yang2012constructing}, browsing hierarchy construction \cite{yang2015browsing}, generating hierarchical summaries \cite{lawrie2003generating}. While most of these techniques work in supervised settings on document collections, our work instead focused on short text queries and offers an unsupervised method of constructing task hierarchies.

Finally, Bayesian Rose Trees and their extensions have been proposed \cite{segal2002probabilistic,blundell2012bayesian,blundell2013bayesian} to model arbitrary branching trees. These algorithms naively cast relationships between objects as binary (0-1) associations while the query-query relationships in general are much richer in content and structure.

We consider a number of such existing methods as baselines and the various advantages of the proposed approach is highlighted in the evaluation section wherein the proposed approach in addition to being more expressive, performs better than state-of-the-art task extraction and hierarchical methods.
\begin{table}[t!]
\centering
\resizebox{!}{!} 
{
\begin{tabular}{ll}
  \textbf{Symbol} & \textbf{Description}\\
  $n_T$ & number of children of tree T\\
  $ab|c$ & partition of set $\lbrace a,b,c\rbrace$ into disjoint sets $\lbrace a,b\rbrace$,$\lbrace c\rbrace$\\
  ch(T) & children of T\\
  $\phi(T)$ & partition of tree T\\
  $p(D_m|T_m)$ & likelihood of data $D_m$ given the tree $T_m$\\
  $\pi_{T_m}$ & mixing proportions of partition of tree $T$\\
  $f(D_m)$ & marginal probability of the data $D_m$\\
  $\mathbb{H}(T)$ & set of all partitions of queries $Q = leaves(T)$\\
  $f(Q)$ & task affinity function for set of queries Q\\
  $r_{q_i,q_j}^k$ & the k-th inter-query affinity between $q_i$ \& $q_j$
\end{tabular}
}
\caption{Table of symbols}
\label{symbols}
\end{table}

\section{Defining Search Tasks}
Jones et al.~\cite{jones2008beyond} was one of the first papers to point out the importance of task representations, where they defined a search task as: 

\theoremstyle{definition}
\begin{definition}{}
A \textit{search task} is an atomic information need resulting in one or more queries.
\end{definition}

Ahmed et al.~\cite{hassan2014supporting} later extended this definition to a more generic one, which can also capture task structures that could possibly consist of related subtasks, each of which could be complex tasks themselves or may finally split down into simpler tasks or atomic informational needs. Following Ahmed \textit{et al.}~\cite{hassan2014supporting}, a complex search task can then be defined as:

\theoremstyle{definition}
\begin{definition}{}
A \textit{complex search task} is a multi-aspect or a multi-step information need consisting of a set of related subtasks, each of which might recursively be complex.
\end{definition}\label{ComplexTask}

The definition of complex tasks is much more generic, and captures all possible search tasks, that can be either complex or atomic (non-complex). Throughout this paper we adopt the definition provided in Definition~\ref{ComplexTask}.2 as the definition for a search task.

Hence, by definition a search task has a hierarchical nature, where each task can consist of an arbitrary number of, possibly complex subtasks. An effective task extraction system should be capable of accurately identifying and representing such hierarchical structures. 

\begin{table*}[t!]
\centering
{
        \begin{tabular}{c|c}
        \hline
        \multicolumn {2}{c}{\textbf{Query-Term Based Affinity ($r^1$)}} \\
        \hline
        cosine & cosine similarity between the term sets of the queries\\
        \hline
       edit & norm edit distance between query strings\\
        \hline
        Jac & Jaccard coeff between the term sets of the queries\\
        \hline
        Term & proportion of common terms between the queries\\
        \hline
        	\multicolumn {2}{c}{\textbf{URL Based Affinity ($r^2$)}} \\
        \hline
        Min-edit-U & Minimum edit distance between all URL pairs from the queries\\
        \hline
        Avg-edit-U & Average edit distance between all URL pairs from the queries\\
        \hline
        Jac-U-min & Minimum Jaccard coefficient between all URL pairs from the queries\\
        \hline
        Jac-U-avg & Average Jaccard coefficient between all URL pairs from the queries\\
        \hline
        \multicolumn {2}{c}{\textbf{Session/User Based Affinity ($r^3$)}} \\
        \hline
        Same-U & if the two queries belong to the same user\\
        \hline
        Same-S & if the two queries belong to the same session\\
        \hline        
        	\multicolumn {2}{c}{\textbf{Embedding Based Affinity ($r^4$)}} \\
        \hline
       Embedding & cosine distance between embedding vectors of the two queries\\
        \hline
        \end{tabular}
}
\caption{Query-Query Affinities.\label{tab:affinity}}
\end{table*}

\section{Constructing Task Hierarchies}
While hierarchical clustering are widely used for clustering, they construct binary trees which may not be the best model to describe data's intrinsic structure in many applications, for example, the task-subtask structure in our case. To remedy this, multi-branch trees are developed. Currently there are few algorithms which generate multi-branch hierarchies. Blundel \textit{et al.}~\cite{blundell2012bayesian,blundell2013bayesian} adopt a simple, deterministic, agglomerative approach called BRTs (Bayesian Rose Trees) for constructing multi-branch hierarchies. In this work, we adapt BRT as a basic algorithm and extend it for constructing task hierarchies. We next describe the major steps of BRT approach.

\begin{figure}[!]
\centering
\includegraphics[width=0.45\textwidth]{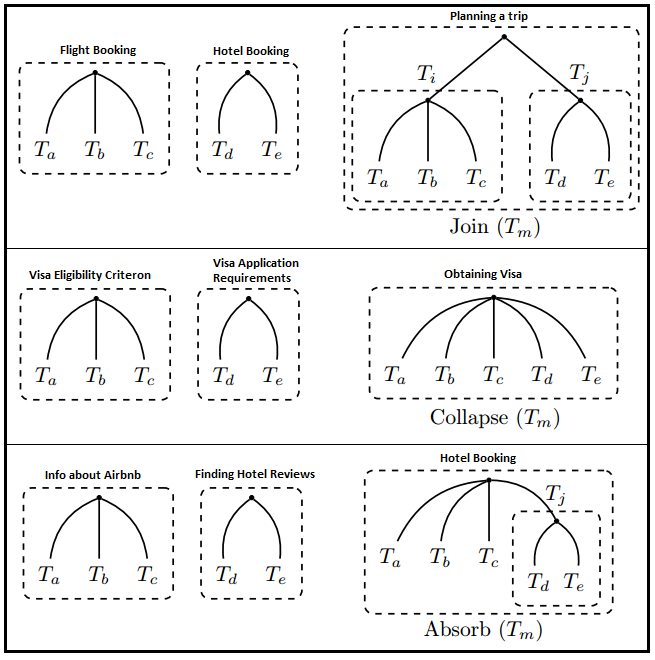}
\vspace{-2mm}
\caption{The different ways of merging trees which allows us to obtain tree structures which best explain the task-subtask structure.}\label{fig4}
\vspace{-2mm}
\end{figure}

\subsection{Bayesian Rose Trees}
BRTs~\cite{blundell2012bayesian,blundell2013bayesian} are based on a greedy probabilistic agglomerative approach to construct multi-branch hierarchies. In the beginning,
each data point is regarded as a tree on its own: $T_i = \{x_i\}$ where $x_i$ is the feature vector of i-th data. For each step, the algorithm selects two trees $T_i$ and $T_j$ and merges them into a new tree $T_m$. Unlike binary hierarchical clustering, BRT uses three possible merging operations, as shown in Figure \ref{fig4}:
\begin{itemize}
\item \textbf{Join}: $T_m = {T_i, T_j}$, such that the tree $T_m$ has two children now
\item \textbf{Absorb}: $T_m = {children(T_i) \cup T_j}$, i.e., the children of one tree gets absorbed into the other tree forming an absorbed tree with $>$2 children
\item \textbf{Collapse}: $T_m = {children(T_i) \cup children(T_j)}$, all the children of both the subtrees get combined together at the same level.
\end{itemize}
Specifically, in each step, the algorithm greedily finds two trees $T_i$ and $T_j$ to merge which maximize the ratio of probability:
\begin{equation}
\frac{p(D_{m}|T_m)}{p(D_{i}|T_i)p(D_{j}|T_j)}
\end{equation}
where $p(D_m|T_m)$ is the likelihood of data $D_m$ given the tree $T_m$, $D_m$ is all the leaf data of $T_m$, and $D_m = D_i \cup D_j$
. The probability $p(D_m|T_m)$ is recursively defined on the children of $T_m$:
\begin{equation}
p(D_m | T_m)  = \pi_{T_m} f(D_m) + (1 - \pi_{T_m}) \prod_{T_i \in ch(T_m)} p(D_i|T_i)
\end{equation}
where $f(D_m)$ is the marginal probability of the data $D_m$ and $\pi_{T_m}$ is the "\textit{mixing proportion}". Intuitively, $\pi_{T_m}$
is the prior probability that all the data in $T_m$ is kept in one cluster instead of partitioned into sub-trees. In BRT\cite{blundell2012bayesian}, $\pi_{T_m}$ is defined as:
\begin{equation}
\pi_{T_m} = 1 - (1 - \gamma)^{n_{T_m}-1}
\end{equation}
where $n_{T_m}$ is the number of children of $T_m$, and $0 \geq \gamma \leq 1$ is the hyperparameter to control the model. A larger $\gamma$ leads to coarser partitions and a smaller $\gamma$ leads to finer partitions. Table \ref{symbols} provides an overview of notations \& symbols used throughout the paper.

\subsection{Building Task Hierarchies}
We next describe our task hierarchy construction approach built on top of Bayesian Rose Trees.  A tree node in our setting is comprised of a group of queries which potentially compose a search task, i.e. these are the set of queries that people tend to issue in order to achieve the task represented in the tree node. 

We define the task-subtask hierarchy recursively: T is a task if either T contains all the queries at its node (an atomic search task) or if T splits into children trees as $T = \{T_1,T_2,...,T_{n_T}\}$ where each of the children trees ($T_i$) are disjoint set of queries corresponding to the $n_T$ subtasks associated with task $T$. This allows us to consider trees as a nested collection of sets of queries defining our task-subtask hierarchical relation.

To form nested hierarchies, we first need to model the query data. This corresponds to defining the marginal distribution of the data $f(D_m)$ as defined in Equation 2. The marginal distribution of the query data ($f(D_m)$) helps us encapsulate insights about task level interdependencies among queries, which aid in constructing better task representations. The original BRT approach \cite{blundell2012bayesian} assumes that the data can be modeled by a set of binary features that follow the Bernoulli distribution. In other words, features (that represent the relationship/similarities between data points) are not weighted and can only be binary. Binary (0/1) relationships are too simplistic to model inter-query relationships; as a result, this major assumption fails to capture the semantic relationships between queries and is not suited for modeling query-task relations. To this end, we propose a novel query affinity model and to alleviate the binary feature assumption imposed by BRT, we propose a conjugate model of query affinities, which we describe next.

\subsection{Conjugate Model of Query Affinities}
\label{taskaffinity}
A tree node in our setting is comprised of a group of queries which \textit{potentially} belong to the same search task. The likelihood of a tree should encapsulate information about the different relationships which exists between queries. Our goal here is to make use of the rich information associated with queries and their result set available to compute the likelihood of a set of queries to belong to the same task. In order to do so, we propose a query affinity model which makes use of a number of different inter-query affinities to determine the tree likelihood function. 

We next describe the technique used to compute four broad categories of inter-query affinity and later describe the Gamma-Poisson conjugate model which makes use of these affinities to compute the marginal distribution of the data. \\

\noindent\textbf{Query-term based Affinity ($r^1$):}\\
Search queries catering to the same or similar informational needs tend to have similar query terms. We make use of this insight and capture query level affinities between a pair of queries. We make use of cosine similarity between the query term sets, the normalized edit distances between queries and the Jaccard Coefficient between query term sets.\\

\noindent\textbf{URL-based Affinity ($r^2$):}\\
Users tackling similar tasks tend to issue queries (possibly different) which return similar URLs, thus encoding the URL level similarity between pairs of queries into the query affinity model helps in capturing another task-specific similarity between queries. Any query pair having high URL level similarity increase the possibility of the query pair originating from similar informational needs. We capture a number of URL-based signals including minimum and average edit distances between URL domains and jaccard coefficient between URLs.\\

\noindent\textbf{User/Session based Affinity ($r^3$):}\\
It is often the case that users issue related queries within a session so as to satisfy their informational need. We leverage this insight by making use of session level information (as a 0/1 binary feature) and user-level information (as a 0/1 binary feature) in our affinity model to identify queries issued in the same session and by the same user accordingly.\\

\noindent\textbf{Query Embedding based Affinity ($r^4$):}\\
Word embeddings capture lexico-semantic regularities in language, such that words with similar syntactic and semantic properties are found to be close to each other in the embedding space. We leverage this insight and propose a query-query affinity metric based on such embeddings. We train a skip-gram word embeddings model where a query term is used as an input to a log-linear classifier with continuous projection layer and words within a certain window before and after the words are predicted. To obtain a query's vector representation, we average the vector representations of each of its query terms and compute the cosine similarity between two queries' vector representations to quantify the embedding based affinity ($r^4$).

Table \ref{tab:affinity} summarizes all features considered to compute these affinities. Our goal is to capture information from all four affinities when defining the likelihood of the tree. We assume that the global affinity among a group of queries can be decomposed into a product of independent terms, each of which represent one of the four affinities from the query-group. For each query group $Q$, we take the normalized sum of the affinities from all pairs of queries in the group $Q$ to form each of the affinity component ($r^k$, k={1,2,3,4}).

Poisson models have been shown as effective query generation models for information retrieval tasks \cite{mei2007study}. While these affinities could be used with a lot of distributions, in the interest of computational efficiency and to avoid approximate solutions, our model will use a hierarchical Gamma-Poisson distribution to encode the query-query affinities. We incorporate the gamma-Poisson conjugate distribution in our model under the assumptions that the query affinities are discretized and for a group of queries $Q$, the affinities can be decomposed to a product of independent terms, each of which represents contributions from the four different affinity types.  Finally, for a tree ($T_m$) consisting of the data ($D_m$), i.e. the set of queries $Q$, we define the marginal likelihood as:
\begin{equation}
f(D_m) = f(Q) = \prod_{k=1}^{k=4} p \bigg( \sum_{i \in 1 \cdots |Q|}\sum_{j \in 1 \cdots |Q|}r^k_{q_i,q_j} | \alpha_k, \beta_k\bigg)
\end{equation}
where $\alpha_k$ \& $\beta_k$ are respectively the shape parameter \& the rate parameter of the four different affinities. Making use of the Poisson-Gamma conjugacy, the probability term in the above product can be written as:
\begin{equation}
p(r|\alpha,\beta) = \int_\lambda p(r|\lambda) p(\lambda|\alpha,\beta)d\lambda
\end{equation}
\begin{equation}
= \Bigg\{ \frac{\Gamma (\alpha +r)}{r!\,\Gamma(\alpha)}\Bigg( \frac{\beta}{\beta +1}\Bigg)^\alpha \Bigg(\frac{1}{\beta +1}\Bigg)^r \Bigg\}
\end{equation}
where $\lambda$ is the Poisson mean rate parameter which gets eliminated from computations because of the Gamma-Poisson conjugacy and where $r$, $\alpha$ \& $\beta$ get replaced by affinity class specific values.

\subsection{Task Coherence based Pruning}
\label{sec:pruning}
The search task extraction algorithm described above provides us a way of constructing a task hierarchy wherein as we go down the tree, nodes comprising of complex multi-aspect tasks split up to provide finer tasks which ideally should model user's fine grained information needs. One key problem with the hierarchy construction algorithm is the continuous splitting of nodes which results in singleton queries occupying the leave nodes. While splitting of nodes which represent complex tasks is important, the nodes representing simple search task queries corresponding to atomic informational needs should not be further split into children nodes. Our goal in this section is to provide a way of quantifying the task complexity of a particular node so as to prevent splitting up nodes representing atomic search task into further subsets of query nodes.

\subsubsection{Identifying Atomic Tasks}
We wish to identify nodes capturing search subtasks which represent atomic informational need. In order to do so, we introduce the notion of \textit{Task Coherence}:

\theoremstyle{definition}
\begin{definition}{}
\textit{Task Coherence} is a measure indicating the atomicity of the information need associated with the task. It is captured by the semantic closeness of the queries associated with the task.
\end{definition}\label{TaskCoherence}

By measuring Task Coherence, we intend to capture the semantic variability of queries within this task in an attempt to identify how complex or atomic a task is. For example, a tree node corresponding to a complex task like planning a vacation would involve queries from varied informational needs including flights, hotels, getaways, etc; while a tree node corresponding to a finer task representing an atomic informational need like finding discount coupons would involve less varied queries - all of which would be about discount coupons. Traditional research in topic modelling has looked into automatic evaluation of topic coherence \cite{newman2010automatic} via Pointwise Mutual Information. We leverage the same insights to capture task coherence.

\subsubsection{Pointwise Mutual Information}
PMI has been studied variously in the context of collocation extraction \cite{pecina2010lexical} and is one measure of the statistical independence of observing two words in close proximity. We wish to compute PMI scores for each node of the tree. A tree node in our discussion so far has been represented by a collection of search queries. We split queries into terms and obtain a set of terms corresponding to each node, and calculate a node's PMI scores using the node's set of query terms.

More specifically, the PMI of a given pair of query terms ($w_1$ \& $w_2$) is given by:
\begin{equation}
PMI(w_1,w_2) = log \frac{p(w_1,w_2)}{p(w_1)p(w_2)}
\end{equation}
where the probabilities are determined from the empirical statistics of some full standard collection. We employ the AOL log query set for this and treat two query terms as co-occurring if both terms occur in the same session. For a given task node ($Q$), we measure task coherence as the average of PMI scores for all pairs of the search terms associated with the task node:
\begin{equation}
PMI-Score(Q) = \frac{1}{|w|} \sum^{|w|}_{i=1} \sum^{|w|}_{j=1} PMI(w_i,w_j)
\end{equation}
where $|w|$ represents the total number of unique search terms associated with task node $Q$. The node's PMI-Score is used as the final measure of task coherence for the task represented via the corresponding node.

\subsubsection{Tree Pruning}
We use the task coherence score associated with each node of the task hierarchy constructed, and prune lower level nodes of the tree to avoid aggressive node splitting. The overall motivation here is to avoid splitting nodes which represent simple search tasks associated with atomic informational needs. We scan through all levels of the search task hierarchy obtained by the algorithm described above and for each node compute its task coherence score. If the task coherence score exceeds a specific threshold, it implies that all the queries in this particular node are aimed at solving the same or very similar informational need and hence, we prune off the sub-tree rooted at this particular node and ignore all further splits of this node. 

\subsection{Algorithmic Overview}
We summarize the overall algorithm to construct the hierarchy by outlining the steps. The problem is treated as one of greedy model selection: each tree T is a different model, and we wish to find the model that best explains the search log data in terms of task-subtask structure.\\

\noindent\textbf{Step 1: Forrest Initialization}:\\
The tree is built in a bottom-up greedy agglomerative fashion, starting from a forest consisting of n (=$|Q|$) trivial trees, each corresponding to exactly one vertex. The algorithm maintains a forest F of trees, the likelihood $p(i) = p(D_{i}|T_i)$ of each tree $T_i \in F$ and the different query affinities. Each iteration then merges two of the trees in the forest. At each iteration, each vertex in the network is a leaf of exactly one tree in the forest. At each iteration a pair of trees in the forest F is chosen to be merged, resulting in forest $F*$.\\

\noindent\textbf{Step 2: Merging Trees}:\\
At each iteration, the best potential merge, say of trees X and Y resulting in tree I, is picked off the heap. Binary trees do not fit into representing search tasks since a task is likely to be composed of more than two subtasks. As a result, following \cite{blundell2013bayesian} we consider three possible mergers of two trees $T_i$ and $T_j$ into $T_m$. $T_m$ may be formed by joining $T_i$ and $T_j$ together using a new node, giving $T_m = \lbrace T_i, T_j\rbrace$. Alternatively $T_m$ may be formed by absorbing $T_i$ as a child of $T_j$, yielding $T_m = \lbrace T_j\rbrace \bigcup ch(T_i)$, or vice-versa, $T_m = \lbrace T_i\rbrace \bigcup ch(T_j)$. We explain the different possible merge operations in Figure \ref{fig4}. We obtain arbitrary shaped sub-trees (without restricting to binary tress) which are better at representing the varied task-subtask structures as observed in search logs with the structures themselves learnt from log data. Such expressive nature of our approach differentiates it from traditional agglomerative clustering approaches which necessarily result in binary trees.\\

\noindent\textbf{Step 3: Model Selection}:\\
Which pair of trees to merge, and how to merge these trees, is determined by considering which pair and type of merger yields the largest Bayes factor improvement over the current model. If the trees $T_i$ and $T_j$ are merged to form the tree M, then the Bayes factor score is:
\begin{equation}
SCORE(M;I,J) = \frac{p(D_{M}|F*)}{p(D_{M}|F)}
\end{equation}
\begin{equation}
 \hspace{4mm} = \frac{p(D_{M}|M)}{p(D_{i}|T_i)p(D_{j}|T_j)}
\end{equation}
where $p(D_{i}|T_i)$ and $p(D_{j}|T_j)$ are given by the dynamic programming equation mentioned above. After a successful merge, the statistics associated with the new tree are updated. Finally, potential mergers of the new tree with other trees in the forest are considered and added onto the heap.

The algorithm finishes when no further merging results in improvement in the Bayes Factor score. Note that the Bayes factor score is based on data local to the merge - i.e., by considering the probability of the connectivity data only among the leaves of the newly merged tree. This permits efficient local computations and makes the assumption that local community structure should depend only on the local connectivity structure.\\

\noindent\textbf{Step 4: Tree Pruning}:\\
After constructing the entire hierarchy, we perform the post-hoc tree pruning procedure described in Section \ref{sec:pruning} wherein we identify atomic task nodes via their task coherence estimates and prune all child nodes of the identified atomic nodes.

\section{Experimental Evaluation}
We perform a number of experiments to evaluate the proposed task-subtask extraction method. First, we compare its performance with existing state-of-the-art task extraction systems on a manually labelled ground-truth dataset and report superior performance (\ref{exp1}). Second, we perform a detailed crowd-sourced evaluation of extracted tasks and additionally validate the hierarchy using human labeled judgments (\ref{exp2}). Third, we show a direct application of the extracted tasks by using the task hierarchy constructed for term prediction  (\ref{exp3}).\\

\noindent\textbf{Parameter Setting:}\\
Unless stated otherwise, we made use of the best performing hyperparameters for the baselines as reported by the authors. The query affinities in the proposed approach were computed from the specific query collection used in the dataset used for each of the three experiments reported below. While hyperparmeter optimization is beyond the scope of this work, we experimented with a range of the shape and inverse scale hyperparameters ($\alpha$, $\beta$) used for the Poison Gamma conjugate model and used the ones which performed best on the validation set for the search task identification results reported in the next section. Additionally, for the tree pruning threshold, we empirically found that a threshold of 0.8 gave the best performance on our toy hierarchies, and was used for all future experiments.

\subsection{Search Task Identification}
\label{exp1}
\noindent To justify the effectiveness of the proposed model in identifying search tasks in query logs, we employ a commonly used AOL data subset with search tasks annotated which is a standard test dataset for evaluating task extraction systems.
We used the task extraction dataset as provided by Lucchese \textit{et al.}\cite{lucchese2011identifying}. The dataset comprises of a sample of 1000 user sessions for which human assessors were asked to manually identify the optimal task-based query sessions, thus producing a ground-truth that can be used for evaluating automatic task-based session discovery methods. For further details on the dataset and the dataset access links, readers are directed to Lucchese \textit{et al.}\cite{lucchese2011identifying}.

We compare our performance with a number of search task identification approaches:
\begin{itemize}
\item \textbf{Bestlink-SVM} \cite{wang2013learning}: This method identified search task using a semi-supervised clustering model based on the latent structural SVM framework.

\item \textbf{QC-HTC/QC-WCC} \cite{lucchese2011identifying}: This series of methods viewed search task identification as the problem of best approximating the manually annotated tasks, and proposed both clustering and heuristic algorithms to solve the problem.

\item \textbf{LDA-Hawkes} \cite{li2014identifying}: a probabilistic method for identifying and labeling search tasks that model query temporal patterns using a special class of point process called Hawkes processes, and combine topic model with Hawkes processes for simultaneously identifying and labeling search tasks.

\item \textbf{LDA Time-Window(TW)}: This model assumes queries belong to the same search task only if they lie in a fixed or flexible time window, and uses LDA to cluster queries into topics based on the query co-occurrences within the same time window. We tested time windows of various sizes and report results on the best performing window size.


\end{itemize}

\subsubsection{Metrics}
A commonly used evaluation metric for search task extraction is the pairwise F-measure computed based on pairwise precision/recall \cite{jones2008beyond,kotov2011modeling} defined as,
\begin{equation}
p_{pair} = \frac{\Sigma_{i\le j} \delta(y(q_i),y(q_j))\delta(\hat{y}(q_i),\hat{y}(q_j))}{\delta(\hat{y}(q_i),\hat{y}(q_j))}
\end{equation}
\begin{equation}
r_{pair} = \frac{\Sigma_{i\le j} \delta(y(q_i),y(q_j))\delta(\hat{y}(q_i),\hat{y}(q_j))}{\delta(y(q_i),y(q_j))}
\end{equation}
where $p_{pair}$ evaluates how many pairs of queries predicted in the same task, i.e., $\delta(\hat{y}(q_i),\hat{y}(q_j) = 1$, are actually annotated as in the same task, i.e., $\delta(y(q_i),y(q_j)) = 1$ and $r_{pair}$ evaluates how many pairs annotated as in the same task are recovered by the algorithm.
Thus, globally F-measure evaluates the extent to which a task contains only queries of a particular annotated task and all queries of that task. Given $p_{pair}$ and $r_{pair}$, the F-measure is computed as:$F_1 = \frac{2\times p_{pair}\times r_{pair}}{p_{pair} + r_{pair}}$.

\subsubsection{Results \& Discussion}
\begin{figure}[t!]
\centering
\includegraphics[width=0.5\textwidth]{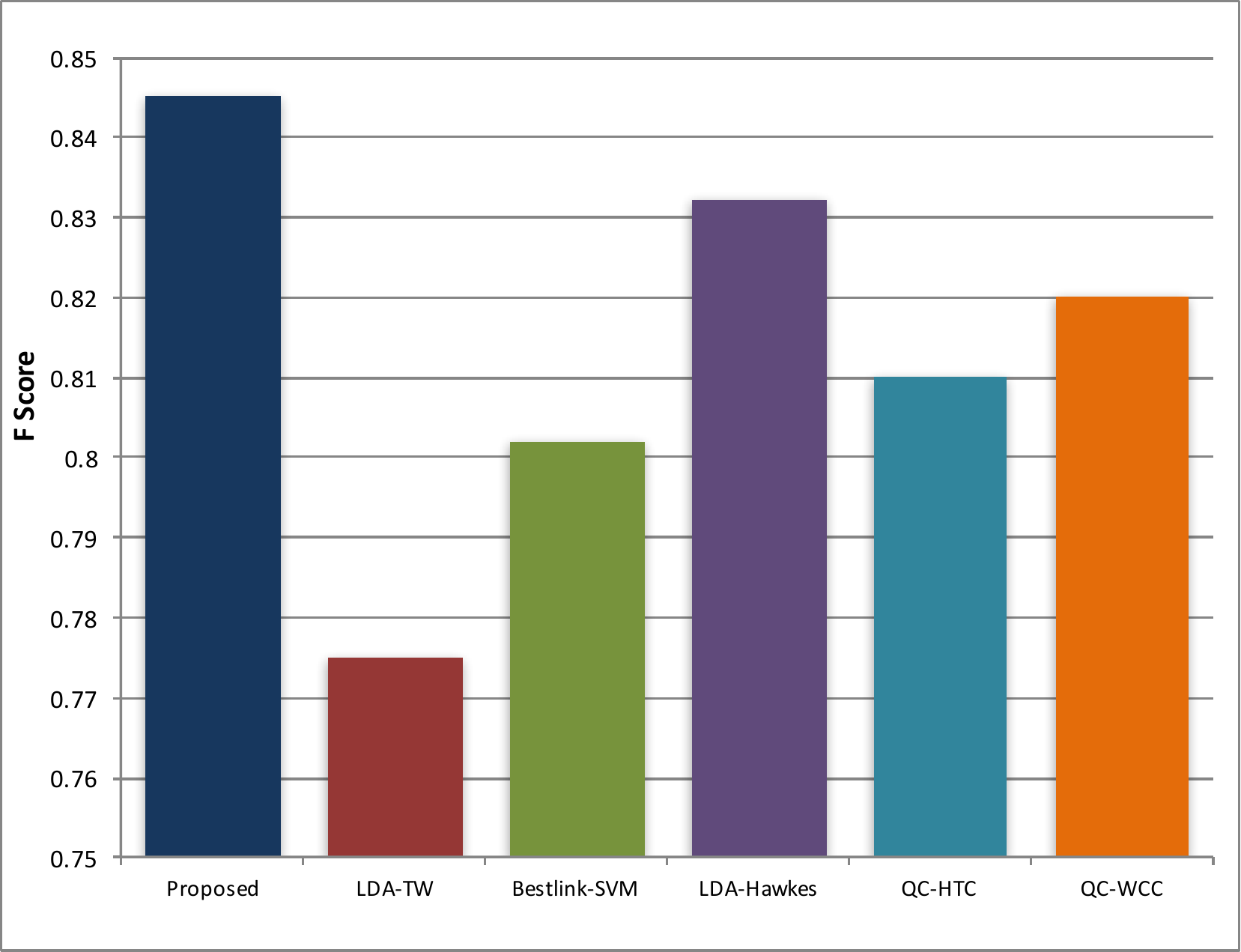}
\caption{F1 score results on AOL tagged dataset}\label{fig5}
\end{figure}

Figure \ref{fig5} compares the proposed model with alternative probabilistic models and state-of-the-art task identification approaches by F1 score. To make fair comparisons, we consider the last level of the pruned tree constructed as task clusters when computing pairwise precision/recall values. It is important to note that the labelled dataset has only flat tasks extracted on a per user basis; as a result, this dataset is not ideal for making fair comparisons of the proposed hierarchy extraction method with baselines. Nevertheless, the proposed approach manages to outperform existing task extraction baselines while having much greater expressive powers and providing the subdivision of tasks into subtasks. LDA-TW performs the worst since its assumptions on query relationship within the same search task are too strong. The advantage over QC-HTC and QC-WCC demonstrates that appropriate usage of query affinity information can even better reflect the semantic relationship between queries, rather than exploiting it in some collaborative knowledge.

\begin{table*}[t]
\centering
\resizebox{0.8\textwidth}{!} 
{
        \begin{tabular}{c|c|c|c|c|c}
        & \multicolumn {4}{c}{\textbf{Task Relatedness}} \\
        \hline
        & \textbf{Proposed}  & \textbf{LDA-TW}  & \textbf{QC-WCC} & \textbf{LDA-Hawkes} & \textbf{QC-HTC}\\
        \hline

       Task Related   & \textbf{72\%*} & 47\% & 60\% & 67\% & 61\%\\
        \hline
        Somewhat Related & 20\% & 14\% & 15\% & 13\%  & 5\% \\
        \hline
        Unrelated & \textbf{10\%} & 23\% & 25\% & 20\% & 34\%\\
        \hline
        \end{tabular}
}

\caption{Performance on Task Relatedness. The results highlighted with * signify statistically significant difference between the proposed approach and best performing baseline using $\chi^2$ test with $p\leq 0.05$.\label{tab:relate}}
\end{table*}
%

\begin{table}[t!]
\centering
\resizebox{0.5\textwidth}{!} 
{
        \begin{tabular}{c|c|c|c|c|c}
        & \multicolumn {4}{c}{\textbf{Subtask Validity}} \\
        \hline
        & \textbf{Proposed}  & \textbf{Jones}  & \textbf{BHCD} & \textbf{BAC}\\
        \hline

        Valid & \textbf{81\%*} & 69\% & 51\%   & 49\%\\
        \hline
        Somewhat Valid &  8\%  & 19\% &  17\% & 21\% \\
        \hline
        Not Valid & 11\% & 12\% & 32\%   & 30\%\\
        \hline
        & \multicolumn {4}{c}{\textbf{Subtask Usefulness}} \\
        \hline
        Useful & \textbf{67\%*} & 52\% & 41\%   & 43\%\\
        \hline
        Somewhat Useful &  8\%  & 17\% & 19\% & 20\%\\
        \hline
        Not Useful & 25\% & 31\%   & 40\% & 37\%\\
        \hline
        \end{tabular}
}

\caption{Performance on Subtask Validity and Subtask Usefulness. Results highlighted with * signify statistically significant difference between the proposed framework and best performing baseline using $\chi^2$ test with $p\leq 0.05$.\label{tab:valid}}
\end{table}

\subsection{Evaluating the Hierarchy}
\label{exp2}
While there are no gold standard datasets for evaluating hierarchies of tasks, we performed crowd-sourced assessments to assess the performance of our hierarchy extraction method. We separately evaluated the coherence and quality of the extracted hierarchies via two different set of judgements obtained via crowdsourcing.\\

\noindent\textbf{\textit{Evaluation Setup}}\\
For the judgment study, we make use of the AOL search logs and randomly sampled entire query history of frequent users who had more than 1000 search queries. The AOL log is a very large and long-term collection consisting of about 20 million of Web queries issued by more than 657000 users over 3 months. We run the task extraction algorithms on the entire set of queries of the sampled users and collect judgments to assess the quality of the tasks extracted. Judgments were provided by over 40 judges who were recruited from the Amazon Mechanical Turk crowdsourcing service. We restricted annotators to those based in the US because the logs came from searchers based in the US. We also used hidden quality control questions to filter out poor-quality judges. The judges were provided with detailed guidelines describing the notion of search tasks and subtasks and were provided with several examples to help them better understand the judgement task.\\

\noindent\textbf{\textit{Evaluating Task Coherence}}\\
In the first study, we evaluated the quality of the tasks extracted by the task extraction algorithms. In an ideal task extraction system, all the queries belonging to the same task cluster should ideally belong to the same task and hence have better task coherence. To this end, we evaluate the task coherence property of the tasks extracted by the different algorithms. For each of the baselines and the proposed algorithm, we select a task at random from the set of tasks extracted and randomly pick up two queries from the selected task. We then ask the human judges the following question:\\

\noindent\textbf{RQ1: Task Relatedness:} Are the given pairs of queries related to the same task? The possible options include (i) Task Related, (ii) Somewhat Task Related and (iii) Unrelated.\\

The task relatedness score provides an estimate of how coherent tasks are. Indeed, a task cluster containing queries from different tasks would score less in Task Relatedness score since if the task cluster is impure, there is a greater chance that the 2 randomly picked queries belong to different tasks and hence get judged Unrelated.\\

\noindent\textbf{\textit{Evaluating the hierarchy}}\\
While there are no gold standard dataset to evaluate hierarchies, in our second crowd-sourced judgment study, we evaluate the quality of the hierarchy extracted. A valid task-subtask hierarchy would have the parent task representing a higher level task with its children tasks representing more focused subtasks, each of which help the user achieve the overall task identified by the parent task.

We evaluate the correctness of the hierarchy by validating parent-child task-subtask relationships. More specifically, we randomly select a parent node from the hierarchy and then randomly select a child node from the set of its immediate child nodes. Given such parent-child node pairs, we randomly pick 5 queries from the parent node and randomly pick 2 queries from the child node. We then show the human judges these parent and child queries and ask the following questions:\\

\noindent\textbf{RQ2: Subtask Validity:} Consider the set of queries representing the search task and the pair of queries representing the subtask. How valid is this subtask given the overall task?\\

The possible judge options include (i) Valid Subtask, (ii) Somewhat valid and (iii) Invalid. Answering this question helps us in analyzing the correctness of the parent-child task-subtask pairs.\\

\noindent\textbf{RQ3: Subtask Usefulness:} Consider the set of queries representing the search task and the pair of queries representing the subtask. Is the subtask useful in completing the overall search task?

The possible judge options include (i) Useful, (ii) Somewhat Useful and (iii) Not Useful. This helps us in evaluating the usefulness of task-subtask pairs by finding the proportion of subtasks which help users in completing the overall task described by the parent node. Overall, the RQ2 and RQ3 help in evaluating the correctness and usefulness of the hierarchy extracted.\\

\noindent\textbf{\textit{Baselines}}\\
Since RQ1 evaluates task coherence without any notion of task-subtask structure, we compare against the top performing baselines from the task extraction setup described in section \ref{exp1}. On the other hand, RQ2 \& RQ3 help in answering questions about the quality of  hierarchy constructed. To make fair comparisons while evaluating the hierarchies, we introduce additional hierarchy extraction baselines:

\begin{itemize}
\item \textbf{Jones Hierarchies} \cite{jones2008beyond}: A supervised learning approach for task boundary detection and same task identification. We train the classifier using the supervised Lucchese AOL task dataset and use it to extract tasks on the current dataset used in the judgment study.

\item \textbf{BHCD} \cite{blundell2013bayesian}: A state-of-the-art bayesian hierarchical community detection algorithm based on stochastic blockmodels and makes use of Beta-Bernoulli conjugate priors to define a network. We build a network of queries and apply BHCD algorithm to extract hierarchies of query communities.

\item \textbf{Bayesian Agglomerative Clustering (BAC)} \cite{heller2005bayesian}: A standard agglomerative hierarchical clustering model based on Dirichlet process mixtures.

\end{itemize}

\noindent\textbf{\textit{Results \& Discussion}}\\
For the first judgment study, each HIT is composed of 20 query pairs per approach being judged for task relatedness. We had three judges work on every HIT. Overall, per method we obtained judgments for 60 query pairs to evaluate the performance on task-relatedness. From among the three judges judging each query-pair, we followed majority voting mechanism to finalize the label for the instance. Table \ref{tab:relate} presents the proportions of query pairs judged as related. About 72\% of query pairs were judged task-related for the proposed approach with LDA-Hawkes performing second best with 67\%. Task relatedness measures how pure the task clusters obtained are, a higher score indicates that the queries belonging to the same task are indeed used for solving the same search task. The overall results indicate that the tasks extracted by the proposed task-subtask extraction algorithm are indeed better than those extracted by the baselines.

For the second judgment study used for evaluating the quality of the hierarchy, we show 10 pairs of parent-child questions in each HIT and ask the human annotators to judge the subtask validity and usefulness. Overall, per method we evaluate 300 such judgments resulting in over 1200 judgments and used maximum voting criterion from among the 3 judges to decide the final label for each instance. Table \ref{tab:valid} compares the performance of the proposed hierarchy extraction method against other hierarchical baselines. The identified subtask was found useful in 67\% cases with the best performing baseline being useful in 52\% of judged instances. This highlights that the extracted hierarchy is indeed composed of better subtasks which are found to be useful in completing the overall task depicted by the parent task. It is interesting to note that for BHCD and BAC baselines, most often the subtasks were found to be invalid and not useful.

Since the same parent-child task-subtask was judged for validity and usefulness, it is expected that the proportion of task-subtasks judged useful would be less than the ones judged valid. Indeed, as can be seen from the Table \ref{tab:valid}, the relative proportions of tasks-subtasks found useful is much less than those found valid.

\begin{figure}[t]
\centering
\includegraphics[width=0.5\textwidth]{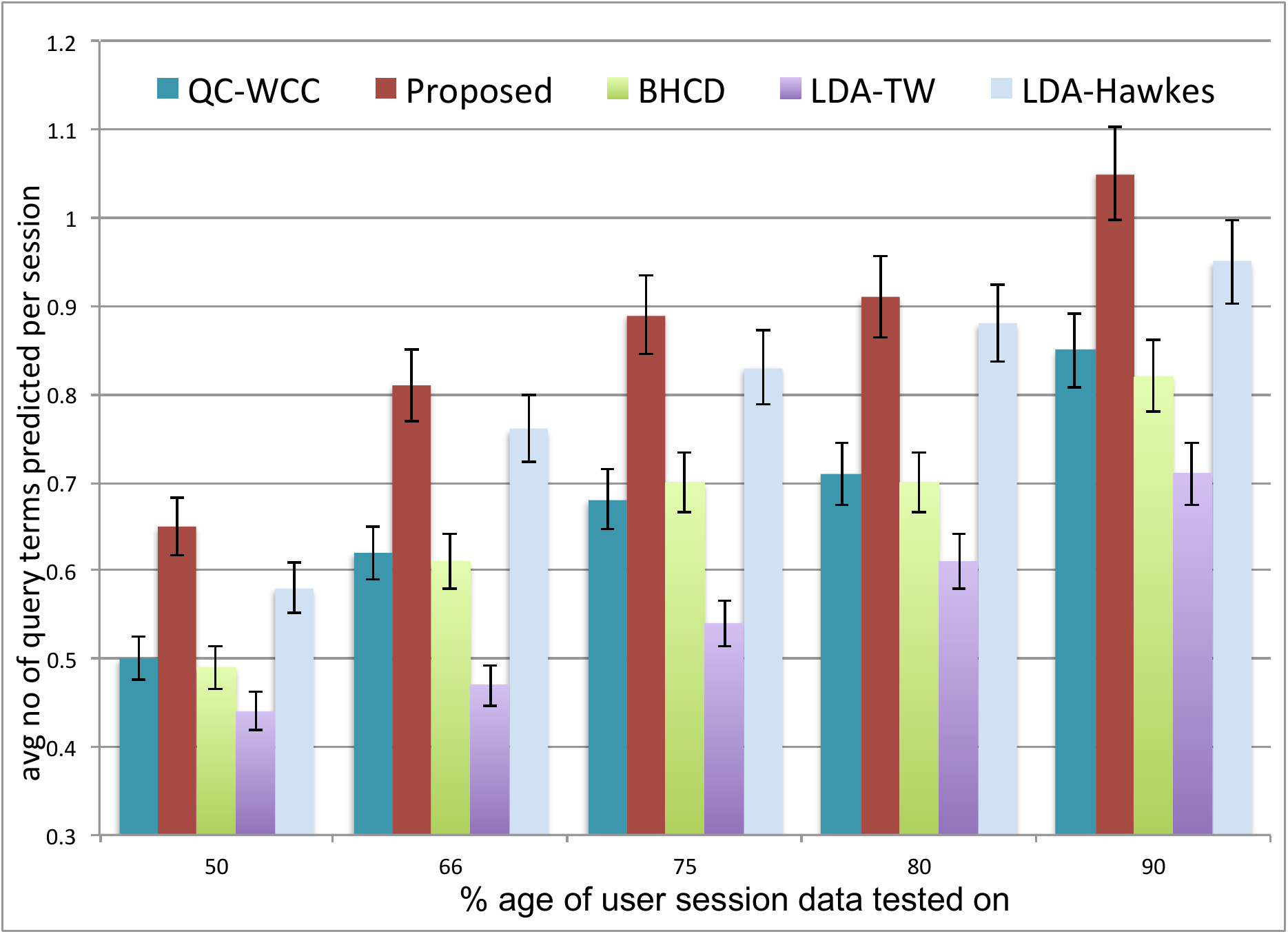}
\caption{Term Prediction performance}\label{fig6}
\vspace{-4mm}
\end{figure}

\subsection{Term Prediction}
\label{exp3}
\noindent In addition to task extraction and user study based evaluation, we chose to follow an indirect evaluation approach based on \textit{Query Term Prediction} wherein given an initial set of queries, we predict future query terms the user may issue later in the session.
This is in line with our goal of supporting users tackling complex search tasks since a task identification system which is capable of identifying "\emph{good}" search tasks will indeed perform better in predicting the set of future query terms.

To evaluate the performance of the proposed task extraction method, we primarily work with the TREC Session Track 2014 \cite{carterette2013overview} and AOL log data and constructed a new dataset consisting of user sessions from AOL logs concerned with Session Track queries. The session track data consists of over 1200 sessions while AOL logs consists of 20M search queries issued by over 657K users. We find the intersection of queries between the Session Track data and AOL logs to identify user sessions in AOL data trying to achieve similar task objectives. The Session Track dataset consists of 60 different \textbf{\textit{topics}}. 
For each of these 60 topics, we separately find user sessions from the entire AOL logs which contain query overlaps with these topics. For each topic, we iterate through the entire AOL logs and select any user session which contains query overlap with the current topic. As a result, we obtain a total of 14030 user sessions which contain around 6.4M queries.

Given the initial queries from a user session and a set of tasks extracted from Session Track data, we leverage queries from the identified task to predict future query terms. For each Session Track topic, we construct a task hierarchy and use the constructed task hierarchy to predict future query terms in the associated user sessions. More specifically, for each topic, we split each user session into two parts: (i) task matching and (ii) held-out evaluation part. We use queries from the task matching part of user sessions to obtain the right node in the task hierarchy from which we then recommend query terms. We pick the tree node which has the highest cosine similarity score based on all the query terms under consideration. We evaluate based on the absolute recall scores - the average number of recommended query terms which match with the query terms in the held-out evaluation part of user sessions.

We baseline against the top performing task extraction baselines from Section \ref{exp1} as well as the top performing hierarchical algorithms from Section \ref{exp2}. To make fair comparisons, we consider nodes at the bottom most level of the pruned tree for task matching and term recommendation.

Figure \ref{fig6} compares the performance on term prediction against the considered baselines. We plot the average number of query terms predicted against the proportion of user session data used. The proposed method is able to better predict future query terms than a standard task extraction baseline as well as a very recent hierarchy construction algorithm.

\section{Conclusion}
Search task hierarchies provide us with a more naturalistic view of considering complex tasks and representing the embedded task-subtask relationships. In this paper we first motivated the need for considering hierarchies of search tasks \& subtasks and presented a novel bayesian nonparametric approach which extracts such hierarchies. We introduced a conjugate query affinity model to capture query affinities to help in task extraction. Finally, we propose the idea of Task Coherence and use it to identify atomic tasks. Our experiments demonstrated the benefits of considering search task hierarchies. Importantly, we were able to demonstrate competitive performance while at the same time outputting a richer and more expressive model of search tasks. This expands the scope for better task recommendation, better search personalization and opens up new avenues for recommendations specifically targeting users based on the tasks they are involved in.

\bibliographystyle{ACM-Reference-Format}
\bibliography{sigproc} 


\begin{thebibliography}{00}


\ifx \showCODEN    \undefined \def \showCODEN     #1{\unskip}     \fi
\ifx \showDOI      \undefined \def \showDOI       #1{{\tt DOI:}\penalty0{#1}\ }
  \fi
\ifx \showISBNx    \undefined \def \showISBNx     #1{\unskip}     \fi
\ifx \showISBNxiii \undefined \def \showISBNxiii  #1{\unskip}     \fi
\ifx \showISSN     \undefined \def \showISSN      #1{\unskip}     \fi
\ifx \showLCCN     \undefined \def \showLCCN      #1{\unskip}     \fi
\ifx \shownote     \undefined \def \shownote      #1{#1}          \fi
\ifx \showarticletitle \undefined \def \showarticletitle #1{#1}   \fi
\ifx \showURL      \undefined \def \showURL       #1{#1}          \fi
\providecommand\bibfield[2]{#2}
\providecommand\bibinfo[2]{#2}
\providecommand\natexlab[1]{#1}
\providecommand\showeprint[2][]{arXiv:#2}

\bibitem[\protect\citeauthoryear{Agichtein, White, Dumais, and
  Bennet}{Agichtein et~al\mbox{.}}{2012}]%
        {agichtein2012search}
\bibfield{author}{\bibinfo{person}{Eugene Agichtein}, \bibinfo{person}{Ryen~W
  White}, \bibinfo{person}{Susan~T Dumais}, {and} \bibinfo{person}{Paul~N
  Bennet}.} \bibinfo{year}{2012}\natexlab{}.
\newblock \showarticletitle{Search, interrupted: understanding and predicting
  search task continuation}. In \bibinfo{booktitle}{{\em Proceedings of the
  35th international ACM SIGIR conference on Research and development in
  information retrieval}}. ACM, \bibinfo{pages}{315--324}.
\newblock


\bibitem[\protect\citeauthoryear{Baeza-Yates, Hurtado, and Mendoza}{Baeza-Yates
  et~al\mbox{.}}{2004}]%
        {baeza2005query}
\bibfield{author}{\bibinfo{person}{Ricardo Baeza-Yates},
  \bibinfo{person}{Carlos Hurtado}, {and} \bibinfo{person}{Marcelo Mendoza}.}
  \bibinfo{year}{2004}\natexlab{}.
\newblock \showarticletitle{Query recommendation using query logs in search
  engines}. In \bibinfo{booktitle}{{\em International Conference on Extending
  Database Technology}}. Springer, \bibinfo{pages}{588--596}.
\newblock


\bibitem[\protect\citeauthoryear{Blundell and Teh}{Blundell and Teh}{2013}]%
        {blundell2013bayesian}
\bibfield{author}{\bibinfo{person}{Charles Blundell} {and}
  \bibinfo{person}{Yee~Whye Teh}.} \bibinfo{year}{2013}\natexlab{}.
\newblock \showarticletitle{Bayesian hierarchical community discovery}. In
  \bibinfo{booktitle}{{\em Advances in Neural Information Processing Systems}}.
  \bibinfo{pages}{1601--1609}.
\newblock


\bibitem[\protect\citeauthoryear{Blundell, Teh, and Heller}{Blundell
  et~al\mbox{.}}{2012}]%
        {blundell2012bayesian}
\bibfield{author}{\bibinfo{person}{Charles Blundell}, \bibinfo{person}{Yee~Whye
  Teh}, {and} \bibinfo{person}{Katherine~A Heller}.}
  \bibinfo{year}{2012}\natexlab{}.
\newblock \showarticletitle{Bayesian rose trees}.
\newblock \bibinfo{journal}{{\em arXiv preprint arXiv:1203.3468\/}}
  (\bibinfo{year}{2012}).
\newblock


\bibitem[\protect\citeauthoryear{Cao, Jiang, Pei, Chen, and Li}{Cao
  et~al\mbox{.}}{2009}]%
        {cao2009towards}
\bibfield{author}{\bibinfo{person}{Huanhuan Cao}, \bibinfo{person}{Daxin
  Jiang}, \bibinfo{person}{Jian Pei}, \bibinfo{person}{Enhong Chen}, {and}
  \bibinfo{person}{Hang Li}.} \bibinfo{year}{2009}\natexlab{}.
\newblock \showarticletitle{Towards context-aware search by learning a very
  large variable length hidden markov model from search logs}. In
  \bibinfo{booktitle}{{\em Proceedings of the 18th international conference on
  World wide web}}. ACM, \bibinfo{pages}{191--200}.
\newblock


\bibitem[\protect\citeauthoryear{Catledge and Pitkow}{Catledge and
  Pitkow}{1995}]%
        {catledge1995characterizing}
\bibfield{author}{\bibinfo{person}{Lara~D Catledge} {and}
  \bibinfo{person}{James~E Pitkow}.} \bibinfo{year}{1995}\natexlab{}.
\newblock \showarticletitle{Characterizing browsing strategies in the
  World-Wide Web}.
\newblock \bibinfo{journal}{{\em Computer Networks and ISDN systems\/}}
  \bibinfo{volume}{27}, \bibinfo{number}{6} (\bibinfo{year}{1995}),
  \bibinfo{pages}{1065--1073}.
\newblock


\bibitem[\protect\citeauthoryear{Chuang and Chien}{Chuang and Chien}{2002}]%
        {chuang2002towards}
\bibfield{author}{\bibinfo{person}{Shui-Lung Chuang} {and}
  \bibinfo{person}{Lee-Feng Chien}.} \bibinfo{year}{2002}\natexlab{}.
\newblock \showarticletitle{Towards automatic generation of query taxonomy: A
  hierarchical query clustering approach}. In \bibinfo{booktitle}{{\em Data
  Mining, 2002. ICDM 2003. Proceedings. 2002 IEEE International Conference
  on}}. IEEE, \bibinfo{pages}{75--82}.
\newblock


\bibitem[\protect\citeauthoryear{Donato, Bonchi, Chi, and Maarek}{Donato
  et~al\mbox{.}}{2010}]%
        {donato2010you}
\bibfield{author}{\bibinfo{person}{Debora Donato}, \bibinfo{person}{Francesco
  Bonchi}, \bibinfo{person}{Tom Chi}, {and} \bibinfo{person}{Yoelle Maarek}.}
  \bibinfo{year}{2010}\natexlab{}.
\newblock \showarticletitle{Do you want to take notes?: identifying research
  missions in Yahoo! search pad}. In \bibinfo{booktitle}{{\em Proceedings of
  the 19th international conference on World wide web}}. ACM,
  \bibinfo{pages}{321--330}.
\newblock


\bibitem[\protect\citeauthoryear{Hassan~Awadallah, White, Pantel, Dumais, and
  Wang}{Hassan~Awadallah et~al\mbox{.}}{2014}]%
        {hassan2014supporting}
\bibfield{author}{\bibinfo{person}{Ahmed Hassan~Awadallah},
  \bibinfo{person}{Ryen~W White}, \bibinfo{person}{Patrick Pantel},
  \bibinfo{person}{Susan~T Dumais}, {and} \bibinfo{person}{Yi-Min Wang}.}
  \bibinfo{year}{2014}\natexlab{}.
\newblock \showarticletitle{Supporting complex search tasks}. In
  \bibinfo{booktitle}{{\em Proceedings of the 23rd ACM International Conference
  on Conference on Information and Knowledge Management}}. ACM,
  \bibinfo{pages}{829--838}.
\newblock


\bibitem[\protect\citeauthoryear{He, G{\"o}ker, and Harper}{He
  et~al\mbox{.}}{2002}]%
        {he2002combining}
\bibfield{author}{\bibinfo{person}{Daqing He}, \bibinfo{person}{Ay{\c{s}}e
  G{\"o}ker}, {and} \bibinfo{person}{David~J Harper}.}
  \bibinfo{year}{2002}\natexlab{}.
\newblock \showarticletitle{Combining evidence for automatic web session
  identification}.
\newblock \bibinfo{journal}{{\em Information Processing \& Management\/}}
  \bibinfo{volume}{38}, \bibinfo{number}{5} (\bibinfo{year}{2002}),
  \bibinfo{pages}{727--742}.
\newblock


\bibitem[\protect\citeauthoryear{Heller and Ghahramani}{Heller and
  Ghahramani}{2005}]%
        {heller2005bayesian}
\bibfield{author}{\bibinfo{person}{Katherine~A Heller} {and}
  \bibinfo{person}{Zoubin Ghahramani}.} \bibinfo{year}{2005}\natexlab{}.
\newblock \showarticletitle{Bayesian hierarchical clustering}. In
  \bibinfo{booktitle}{{\em Proceedings of the 22nd international conference on
  Machine learning}}. ACM, \bibinfo{pages}{297--304}.
\newblock


\bibitem[\protect\citeauthoryear{Hua, Song, Wang, and Zhou}{Hua
  et~al\mbox{.}}{2013}]%
        {hua2013identifying}
\bibfield{author}{\bibinfo{person}{Wen Hua}, \bibinfo{person}{Yangqiu Song},
  \bibinfo{person}{Haixun Wang}, {and} \bibinfo{person}{Xiaofang Zhou}.}
  \bibinfo{year}{2013}\natexlab{}.
\newblock \showarticletitle{Identifying users' topical tasks in web search}. In
  \bibinfo{booktitle}{{\em Proceedings of the sixth ACM international
  conference on Web search and data mining}}. ACM, \bibinfo{pages}{93--102}.
\newblock


\bibitem[\protect\citeauthoryear{Jones and Klinkner}{Jones and
  Klinkner}{2008}]%
        {jones2008beyond}
\bibfield{author}{\bibinfo{person}{Rosie Jones} {and}
  \bibinfo{person}{Kristina~Lisa Klinkner}.} \bibinfo{year}{2008}\natexlab{}.
\newblock \showarticletitle{Beyond the session timeout: automatic hierarchical
  segmentation of search topics in query logs}. In \bibinfo{booktitle}{{\em
  Proceedings of the 17th ACM conference on Information and knowledge
  management}}. ACM, \bibinfo{pages}{699--708}.
\newblock


\bibitem[\protect\citeauthoryear{Jones, Rey, Madani, and Greiner}{Jones
  et~al\mbox{.}}{2006}]%
        {jones2006generating}
\bibfield{author}{\bibinfo{person}{Rosie Jones}, \bibinfo{person}{Benjamin
  Rey}, \bibinfo{person}{Omid Madani}, {and} \bibinfo{person}{Wiley Greiner}.}
  \bibinfo{year}{2006}\natexlab{}.
\newblock \showarticletitle{Generating query substitutions}. In
  \bibinfo{booktitle}{{\em Proceedings of the 15th international conference on
  World Wide Web}}. ACM, \bibinfo{pages}{387--396}.
\newblock


\bibitem[\protect\citeauthoryear{Kotov, Bennett, White, Dumais, and
  Teevan}{Kotov et~al\mbox{.}}{2011}]%
        {kotov2011modeling}
\bibfield{author}{\bibinfo{person}{Alexander Kotov}, \bibinfo{person}{Paul~N
  Bennett}, \bibinfo{person}{Ryen~W White}, \bibinfo{person}{Susan~T Dumais},
  {and} \bibinfo{person}{Jaime Teevan}.} \bibinfo{year}{2011}\natexlab{}.
\newblock \showarticletitle{Modeling and analysis of cross-session search
  tasks}. In \bibinfo{booktitle}{{\em Proceedings of the 34th international ACM
  SIGIR conference on Research and development in Information Retrieval}}. ACM,
  \bibinfo{pages}{5--14}.
\newblock


\bibitem[\protect\citeauthoryear{Lawrie and Croft}{Lawrie and Croft}{2003}]%
        {lawrie2003generating}
\bibfield{author}{\bibinfo{person}{Dawn~J Lawrie} {and}
  \bibinfo{person}{W~Bruce Croft}.} \bibinfo{year}{2003}\natexlab{}.
\newblock \showarticletitle{Generating hierarchical summaries for web
  searches}. In \bibinfo{booktitle}{{\em Proceedings of the 26th annual
  international ACM SIGIR conference on Research and development in informaion
  retrieval}}. ACM, \bibinfo{pages}{457--458}.
\newblock


\bibitem[\protect\citeauthoryear{Li, Deng, Dong, Chang, and Zha}{Li
  et~al\mbox{.}}{2014}]%
        {li2014identifying}
\bibfield{author}{\bibinfo{person}{Liangda Li}, \bibinfo{person}{Hongbo Deng},
  \bibinfo{person}{Anlei Dong}, \bibinfo{person}{Yi Chang}, {and}
  \bibinfo{person}{Hongyuan Zha}.} \bibinfo{year}{2014}\natexlab{}.
\newblock \showarticletitle{Identifying and labeling search tasks via
  query-based hawkes processes}. In \bibinfo{booktitle}{{\em Proceedings of the
  20th ACM SIGKDD international conference on Knowledge discovery and data
  mining}}. ACM, \bibinfo{pages}{731--740}.
\newblock


\bibitem[\protect\citeauthoryear{Liao, Song, He, and Huang}{Liao
  et~al\mbox{.}}{2012}]%
        {liao2012evaluating}
\bibfield{author}{\bibinfo{person}{Zhen Liao}, \bibinfo{person}{Yang Song},
  \bibinfo{person}{Li-wei He}, {and} \bibinfo{person}{Yalou Huang}.}
  \bibinfo{year}{2012}\natexlab{}.
\newblock \showarticletitle{Evaluating the effectiveness of search task
  trails}. In \bibinfo{booktitle}{{\em Proceedings of the 21st international
  conference on World Wide Web}}. ACM, \bibinfo{pages}{489--498}.
\newblock


\bibitem[\protect\citeauthoryear{Liu, Song, Liu, and Wang}{Liu
  et~al\mbox{.}}{2012}]%
        {liu2012automatic}
\bibfield{author}{\bibinfo{person}{Xueqing Liu}, \bibinfo{person}{Yangqiu
  Song}, \bibinfo{person}{Shixia Liu}, {and} \bibinfo{person}{Haixun Wang}.}
  \bibinfo{year}{2012}\natexlab{}.
\newblock \showarticletitle{Automatic taxonomy construction from keywords}. In
  \bibinfo{booktitle}{{\em Proceedings of the 18th ACM SIGKDD international
  conference on Knowledge discovery and data mining}}. ACM,
  \bibinfo{pages}{1433--1441}.
\newblock


\bibitem[\protect\citeauthoryear{Lucchese, Orlando, Perego, Silvestri, and
  Tolomei}{Lucchese et~al\mbox{.}}{2011}]%
        {lucchese2011identifying}
\bibfield{author}{\bibinfo{person}{Claudio Lucchese},
  \bibinfo{person}{Salvatore Orlando}, \bibinfo{person}{Raffaele Perego},
  \bibinfo{person}{Fabrizio Silvestri}, {and} \bibinfo{person}{Gabriele
  Tolomei}.} \bibinfo{year}{2011}\natexlab{}.
\newblock \showarticletitle{Identifying task-based sessions in search engine
  query logs}. In \bibinfo{booktitle}{{\em Proceedings of the fourth ACM
  international conference on Web search and data mining}}. ACM,
  \bibinfo{pages}{277--286}.
\newblock


\bibitem[\protect\citeauthoryear{Lucchese, Orlando, Perego, Silvestri, and
  Tolomei}{Lucchese et~al\mbox{.}}{2013}]%
        {lucchese2013discovering}
\bibfield{author}{\bibinfo{person}{Claudio Lucchese},
  \bibinfo{person}{Salvatore Orlando}, \bibinfo{person}{Raffaele Perego},
  \bibinfo{person}{Fabrizio Silvestri}, {and} \bibinfo{person}{Gabriele
  Tolomei}.} \bibinfo{year}{2013}\natexlab{}.
\newblock \showarticletitle{Discovering tasks from search engine query logs}.
\newblock \bibinfo{journal}{{\em ACM Transactions on Information Systems
  (TOIS)\/}} \bibinfo{volume}{31}, \bibinfo{number}{3} (\bibinfo{year}{2013}),
  \bibinfo{pages}{14}.
\newblock


\bibitem[\protect\citeauthoryear{Ma~Kay and Watters}{Ma~Kay and
  Watters}{2008}]%
        {ma2008exploring}
\bibfield{author}{\bibinfo{person}{Bonnie Ma~Kay} {and}
  \bibinfo{person}{Carolyn Watters}.} \bibinfo{year}{2008}\natexlab{}.
\newblock \showarticletitle{Exploring multi-session web tasks}. In
  \bibinfo{booktitle}{{\em Proceedings of the SIGCHI Conference on Human
  Factors in Computing Systems}}. ACM, \bibinfo{pages}{1187--1196}.
\newblock


\bibitem[\protect\citeauthoryear{Mehrotra, Bhattacharya, and Yilmaz}{Mehrotra
  et~al\mbox{.}}{2016}]%
        {mehrotra2016characterizing}
\bibfield{author}{\bibinfo{person}{Rishabh Mehrotra}, \bibinfo{person}{Prasanta
  Bhattacharya}, {and} \bibinfo{person}{Emine Yilmaz}.}
  \bibinfo{year}{2016}\natexlab{}.
\newblock \showarticletitle{Characterizing users' multi-tasking behavior in web
  search}. In \bibinfo{booktitle}{{\em Proceedings of the 2016 ACM on
  Conference on Human Information Interaction and Retrieval}}. ACM,
  \bibinfo{pages}{297--300}.
\newblock


\bibitem[\protect\citeauthoryear{Mehrotra and Yilmaz}{Mehrotra and
  Yilmaz}{2015a}]%
        {mehrotra2015terms}
\bibfield{author}{\bibinfo{person}{Rishabh Mehrotra} {and}
  \bibinfo{person}{Emine Yilmaz}.} \bibinfo{year}{2015}\natexlab{a}.
\newblock \showarticletitle{Terms, topics \& tasks: Enhanced user modelling for
  better personalization}. In \bibinfo{booktitle}{{\em Proceedings of the 2015
  International Conference on The Theory of Information Retrieval}}. ACM,
  \bibinfo{pages}{131--140}.
\newblock


\bibitem[\protect\citeauthoryear{Mehrotra and Yilmaz}{Mehrotra and
  Yilmaz}{2015b}]%
        {mehrotra2015towards}
\bibfield{author}{\bibinfo{person}{Rishabh Mehrotra} {and}
  \bibinfo{person}{Emine Yilmaz}.} \bibinfo{year}{2015}\natexlab{b}.
\newblock \showarticletitle{Towards hierarchies of search tasks \& subtasks}.
  In \bibinfo{booktitle}{{\em Proceedings of the 24th International Conference
  on World Wide Web}}. ACM, \bibinfo{pages}{73--74}.
\newblock


\bibitem[\protect\citeauthoryear{Mei, Fang, and Zhai}{Mei
  et~al\mbox{.}}{2007}]%
        {mei2007study}
\bibfield{author}{\bibinfo{person}{Qiaozhu Mei}, \bibinfo{person}{Hui Fang},
  {and} \bibinfo{person}{ChengXiang Zhai}.} \bibinfo{year}{2007}\natexlab{}.
\newblock \showarticletitle{A study of Poisson query generation model for
  information retrieval}. In \bibinfo{booktitle}{{\em Proceedings of the 30th
  annual international ACM SIGIR conference on Research and development in
  information retrieval}}. ACM, \bibinfo{pages}{319--326}.
\newblock


\bibitem[\protect\citeauthoryear{Mei, Zhou, and Church}{Mei
  et~al\mbox{.}}{2008}]%
        {mei2008query}
\bibfield{author}{\bibinfo{person}{Qiaozhu Mei}, \bibinfo{person}{Dengyong
  Zhou}, {and} \bibinfo{person}{Kenneth Church}.}
  \bibinfo{year}{2008}\natexlab{}.
\newblock \showarticletitle{Query suggestion using hitting time}. In
  \bibinfo{booktitle}{{\em Proceedings of the 17th ACM conference on
  Information and knowledge management}}. ACM, \bibinfo{pages}{469--478}.
\newblock


\bibitem[\protect\citeauthoryear{Morris, Ringel~Morris, and Venolia}{Morris
  et~al\mbox{.}}{2008}]%
        {morris2008searchbar}
\bibfield{author}{\bibinfo{person}{Dan Morris}, \bibinfo{person}{Meredith
  Ringel~Morris}, {and} \bibinfo{person}{Gina Venolia}.}
  \bibinfo{year}{2008}\natexlab{}.
\newblock \showarticletitle{SearchBar: a search-centric web history for task
  resumption and information re-finding}. In \bibinfo{booktitle}{{\em
  Proceedings of the SIGCHI Conference on Human Factors in Computing Systems}}.
  ACM, \bibinfo{pages}{1207--1216}.
\newblock


\bibitem[\protect\citeauthoryear{Newman, Lau, Grieser, and Baldwin}{Newman
  et~al\mbox{.}}{2010}]%
        {newman2010automatic}
\bibfield{author}{\bibinfo{person}{David Newman}, \bibinfo{person}{Jey~Han
  Lau}, \bibinfo{person}{Karl Grieser}, {and} \bibinfo{person}{Timothy
  Baldwin}.} \bibinfo{year}{2010}\natexlab{}.
\newblock \showarticletitle{Automatic evaluation of topic coherence}. In
  \bibinfo{booktitle}{{\em Human Language Technologies: The 2010 Annual
  Conference of the North American Chapter of the Association for Computational
  Linguistics}}. Association for Computational Linguistics,
  \bibinfo{pages}{100--108}.
\newblock


\bibitem[\protect\citeauthoryear{O'Connor, Krieger, and Ahn}{O'Connor
  et~al\mbox{.}}{2010}]%
        {o2010tweetmotif}
\bibfield{author}{\bibinfo{person}{Brendan O'Connor}, \bibinfo{person}{Michel
  Krieger}, {and} \bibinfo{person}{David Ahn}.}
  \bibinfo{year}{2010}\natexlab{}.
\newblock \showarticletitle{TweetMotif: Exploratory Search and Topic
  Summarization for Twitter.}. In \bibinfo{booktitle}{{\em ICWSM}}.
  \bibinfo{pages}{384--385}.
\newblock


\bibitem[\protect\citeauthoryear{Pecina}{Pecina}{2010}]%
        {pecina2010lexical}
\bibfield{author}{\bibinfo{person}{Pavel Pecina}.}
  \bibinfo{year}{2010}\natexlab{}.
\newblock \showarticletitle{Lexical association measures and collocation
  extraction}.
\newblock \bibinfo{journal}{{\em Language resources and evaluation\/}}
  \bibinfo{volume}{44}, \bibinfo{number}{1-2} (\bibinfo{year}{2010}),
  \bibinfo{pages}{137--158}.
\newblock


\bibitem[\protect\citeauthoryear{Segal and Koller}{Segal and Koller}{2002}]%
        {segal2002probabilistic}
\bibfield{author}{\bibinfo{person}{Eran Segal} {and} \bibinfo{person}{Daphne
  Koller}.} \bibinfo{year}{2002}\natexlab{}.
\newblock \showarticletitle{Probabilistic hierarchical clustering for
  biological data}. In \bibinfo{booktitle}{{\em Proceedings of the sixth annual
  international conference on Computational biology}}. ACM,
  \bibinfo{pages}{273--280}.
\newblock


\bibitem[\protect\citeauthoryear{Silverstein, Marais, Henzinger, and
  Moricz}{Silverstein et~al\mbox{.}}{1999}]%
        {silverstein1999analysis}
\bibfield{author}{\bibinfo{person}{Craig Silverstein}, \bibinfo{person}{Hannes
  Marais}, \bibinfo{person}{Monika Henzinger}, {and} \bibinfo{person}{Michael
  Moricz}.} \bibinfo{year}{1999}\natexlab{}.
\newblock \showarticletitle{Analysis of a very large web search engine query
  log}. In \bibinfo{booktitle}{{\em ACm SIGIR Forum}},
  Vol.~\bibinfo{volume}{33}. ACM, \bibinfo{pages}{6--12}.
\newblock


\bibitem[\protect\citeauthoryear{Singla, White, and Huang}{Singla
  et~al\mbox{.}}{2010}]%
        {singla2010studying}
\bibfield{author}{\bibinfo{person}{Adish Singla}, \bibinfo{person}{Ryen White},
  {and} \bibinfo{person}{Jeff Huang}.} \bibinfo{year}{2010}\natexlab{}.
\newblock \showarticletitle{Studying trailfinding algorithms for enhanced web
  search}. In \bibinfo{booktitle}{{\em Proceedings of the 33rd international
  ACM SIGIR conference on Research and development in information retrieval}}.
  ACM, \bibinfo{pages}{443--450}.
\newblock


\bibitem[\protect\citeauthoryear{Spink, Koshman, Park, Field, and Jansen}{Spink
  et~al\mbox{.}}{2005}]%
        {spink2005multitasking}
\bibfield{author}{\bibinfo{person}{Amanda Spink}, \bibinfo{person}{Sherry
  Koshman}, \bibinfo{person}{Minsoo Park}, \bibinfo{person}{Chris Field}, {and}
  \bibinfo{person}{Bernard~J Jansen}.} \bibinfo{year}{2005}\natexlab{}.
\newblock \showarticletitle{Multitasking web search on vivisimo. com}. In
  \bibinfo{booktitle}{{\em Information Technology: Coding and Computing, 2005.
  ITCC 2005. International Conference on}}, Vol.~\bibinfo{volume}{2}. IEEE,
  \bibinfo{pages}{486--490}.
\newblock


\bibitem[\protect\citeauthoryear{Wang, Song, Chang, He, Hassan, and White}{Wang
  et~al\mbox{.}}{2014}]%
        {wang2014modeling}
\bibfield{author}{\bibinfo{person}{Hongning Wang}, \bibinfo{person}{Yang Song},
  \bibinfo{person}{Ming-Wei Chang}, \bibinfo{person}{Xiaodong He},
  \bibinfo{person}{Ahmed Hassan}, {and} \bibinfo{person}{Ryen~W White}.}
  \bibinfo{year}{2014}\natexlab{}.
\newblock \showarticletitle{Modeling action-level satisfaction for search task
  satisfaction prediction}. In \bibinfo{booktitle}{{\em Proceedings of the 37th
  international ACM SIGIR conference on Research \& development in information
  retrieval}}. ACM, \bibinfo{pages}{123--132}.
\newblock


\bibitem[\protect\citeauthoryear{Wang, Song, Chang, He, White, and Chu}{Wang
  et~al\mbox{.}}{2013}]%
        {wang2013learning}
\bibfield{author}{\bibinfo{person}{Hongning Wang}, \bibinfo{person}{Yang Song},
  \bibinfo{person}{Ming-Wei Chang}, \bibinfo{person}{Xiaodong He},
  \bibinfo{person}{Ryen~W White}, {and} \bibinfo{person}{Wei Chu}.}
  \bibinfo{year}{2013}\natexlab{}.
\newblock \showarticletitle{Learning to extract cross-session search tasks}. In
  \bibinfo{booktitle}{{\em Proceedings of the 22nd international conference on
  World Wide Web}}. ACM, \bibinfo{pages}{1353--1364}.
\newblock


\bibitem[\protect\citeauthoryear{White, Chu, Hassan, He, Song, and Wang}{White
  et~al\mbox{.}}{2013}]%
        {white2013enhancing}
\bibfield{author}{\bibinfo{person}{Ryen~W White}, \bibinfo{person}{Wei Chu},
  \bibinfo{person}{Ahmed Hassan}, \bibinfo{person}{Xiaodong He},
  \bibinfo{person}{Yang Song}, {and} \bibinfo{person}{Hongning Wang}.}
  \bibinfo{year}{2013}\natexlab{}.
\newblock \showarticletitle{Enhancing personalized search by mining and
  modeling task behavior}. In \bibinfo{booktitle}{{\em Proceedings of the 22nd
  international conference on World Wide Web}}. ACM,
  \bibinfo{pages}{1411--1420}.
\newblock


\bibitem[\protect\citeauthoryear{Yang}{Yang}{2012}]%
        {yang2012constructing}
\bibfield{author}{\bibinfo{person}{Hui Yang}.} \bibinfo{year}{2012}\natexlab{}.
\newblock \showarticletitle{Constructing task-specific taxonomies for document
  collection browsing}. In \bibinfo{booktitle}{{\em Proceedings of the 2012
  Joint Conference on Empirical Methods in Natural Language Processing and
  Computational Natural Language Learning}}. Association for Computational
  Linguistics, \bibinfo{pages}{1278--1289}.
\newblock


\bibitem[\protect\citeauthoryear{Yang}{Yang}{2015}]%
        {yang2015browsing}
\bibfield{author}{\bibinfo{person}{Hui Yang}.} \bibinfo{year}{2015}\natexlab{}.
\newblock \showarticletitle{Browsing hierarchy construction by minimum
  evolution}.
\newblock \bibinfo{journal}{{\em ACM Transactions on Information Systems
  (TOIS)\/}} \bibinfo{volume}{33}, \bibinfo{number}{3} (\bibinfo{year}{2015}),
  \bibinfo{pages}{13}.
\newblock


\bibitem[\protect\citeauthoryear{Zhang, Zhang, Liu, Tat-Seng, Zhang, and
  Ma}{Zhang et~al\mbox{.}}{2015}]%
        {zhang2015task}
\bibfield{author}{\bibinfo{person}{Yongfeng Zhang}, \bibinfo{person}{Min
  Zhang}, \bibinfo{person}{Yiqun Liu}, \bibinfo{person}{Chua Tat-Seng},
  \bibinfo{person}{Yi Zhang}, {and} \bibinfo{person}{Shaoping Ma}.}
  \bibinfo{year}{2015}\natexlab{}.
\newblock \showarticletitle{Task-based recommendation on a web-scale}. In
  \bibinfo{booktitle}{{\em Big Data (Big Data), 2015 IEEE International
  Conference on}}. IEEE, \bibinfo{pages}{827--836}.
\newblock


\end{thebibliography}

\end{document}